# Content Delivery Through Hybrid Architecture in Video on Demand System

Soumen Kanrar[1,2*], Soamdeep Singha[2]

[1] Department of Computer Science & Engineering, DIT University, Dehradun Uttarakhand 248009, India
[2] Department of Computer Science, Vidyasagar University, Midnapore West Bengal 721102, India

Corresponding Author Email: Soumen.kanrar@dituniversity.edu.in

**ABSTRACT**

Peer-to-Peer (P2P) network needs architectural modification for smooth and fast transportation of video content. The viewer imports chunk video objects through the proxy server. The enormous growth of user requests in a small session of time creates huge load on the VOD system. The situation requires either the proxy server streamed video-content fully or partly to the viewers. The missing chunk at the proxy server is imported from the connected peer nodes. Peers exchange chunks among themselves according to some chunk selection policy. Peer node randomly contacts another peer to download a missing chunk from the buffers during each time slot. In video streaming, when the relevant frame is required at the viewer ends that should be available at the respective proxy server. The video watcher also initiates various types of interactive operations like a move forward or skips some finite number of frames that create congestion inside the VOD system. To elevate the situation it needs an effective content delivery mechanism for smooth transportation of content. The proposed hybrid architecture is composed of P2P and mesh architecture that effectively enhances the search mechanism and content transportation in the VOD system.

## 1. INTRODUCTION

Visual multimedia source combines a sequence of images to form a moving picture. The video transmits signals to a screen and processes the order in which the screen captures the show. Videos usually have audio components that correspond with the pictures being shown on the screen. In general, a video is a sequence of images processed electronically into an analog or digital format and it is played on a screen with sufficient rapidity as to create the illusion of motion and continuity. The compressing technique is used to compress each individual still image from the collected set of images. Multimedia is the merging of major components such as computation and communication. Researchers are contributing to multimedia computation into the multimedia storage server and associated software development tools. The notable contributions are found on the domain, such as music computation, computer–aided learning, interactive video streaming, dynamic adaptive prefetching [1-3]. The distributed system enhances the video content delivery mechanism with the combination of multimedia computing [4-5]. The primitive characteristic of multimedia systems is the incorporation of continuous media such as voice, video, and animation. The distributed multimedia system requires a continuous flow of data for long periods of time. In this context, it should be considered, play out of video streams captured by the remote camera as a part of direct streaming [6]. Here, the encoded and compressed data i.e., content is transported over the heterogeneous network. Generally, the encoding takes more time to compare with decoding. There are some videos, which are pre-encoded in advance. So, it is required that the decoding latency should be reasonably low during streaming to viewer side application. Each video frame is partitioned into a collection of blocks. The blocks are analyzed by the encoder to determine the correct blocks, before transmission. The blocks contain a significant amount of changes from frame to frame. Each block is predicted by means of temporal spatiality variation. The spatial prediction or inter-frame prediction is used by a block identifier for a group of samples. These provide all the similar types characteristic like intensity, color for every individual frame etc. The identifier is transported for decoding purpose. The temporal prediction predicts the inter-frame motion that compensates, any existence of inaccuracy. This is a prediction error or as usual a residual. The residual is transformed into compact form as a block by block basis. The residual and predicted information like the prediction about the mode and motion vectors (MVs). Those are forwarded together for decoding purpose, as to reconstruct the original video frame. The bit rate adaption or transcoding converts the high-bitrate streams into a low-bit stream. The transcoder is used to process and converting one bitstream to another bitstream. A video transcoder can perform several functions such as bit rate adjustment, spatial or temporal adjustment. The conversion is used to transmit bit streams of the content over the connected heterogeneous networks where the bandwidths of the transmission paths are not identical for each link. The buffering mechanism makes smooth the video traffic irregularity. But the quality of service for the timely delivery of video content becomes an issue. If the bit rate is decreased then the adaptive visual quality is also reduced. It is often needed to adopt the bit-rate for the coded content, the bit-stream to the available bandwidth helps to spread over the heterogeneous network environments. The issue remains a challenge to deliver the coded content to the viewer's buffer as quickly as possible by enhancing the architecture, protocol, blockage control at junction point and algorithm. Aiming towards that direction some of the referenced works on blockage measurement during the content delivery without MPLS (Multiprotocol Label Switching) overlay [7]. Ning Wang et.al, have measured the traffic blockage in the network without considering the switching mechanism. Chao Liang et. al., have presented the Chunk delay during the video streaming over Peer-to-Peer (P2P) network, by average hop counts [8]. C. A. V. Melo et al., addressed the issue related to P2P overlay that assists CDNs on duties of distributing short videos [9]. Yingnan Zhu et al., have measured the delay for a concurrent session for video streaming over wireless mesh architecture [10]. Kideok Cho et al., Zhou Su et. al., and Muaz Niazi, have shown the best way to present the traffic scenario and delay in the network during video streaming by considering hop count [11-13]. In this context, the characterizing churn requires adequate unbiased information about the arrival and departure of peer nodes in the P2P network [14]. Its need to captures the heterogeneous behavior of the peers. Dynamic adaptive prefetching with the transcoding is the trends in cloud-based streaming [1]. The hybrid architecture helps to analyze link dynamics and the ability of the Video on Demand system to stay connected under churn. The P2P networks broadly classified as unstructured and structured networks [15]. The unstructured networks organize peers onto random graphs; onwards the graphs are constructed based on some pre-defined rule. The links between the nodes share common structured patterns. The main advantage of unstructured P2P streaming over structured P2P streaming is that the unstructured approach has robust performance under peer churn [16]. Negin Golrezaei et al., have proposed the concept of distributed storage for video content and popularity distributions for a small size campus network during the searching of video content in an unstructured network [17]. Golrezaei et al., have considered an empirical distribution of video popularity from a wired network, namely the campus network of the University of Massachusetts at Amherst. To unraveling the impact of temporal and geographical locality in a content transfer, Stefano Traverso et al., have evaluated the volume of user requests over three weeks for video traffic trace data [18]. Konstantinos Poularakis et al., have explored the multicast video data content for wireless networks [19]. Cheng Zhan et al., have exhibited the ability to alleviate the pressure brought by the explosion of video traffic on the present architecture of cellular networks and requires the smooth content transfer [20-21]. This paper is organized as follows. Section 1 introduces the problem. The basic, video content encryption mechanisms and compression technique have been discussed in section 2, followed with transcoding for video content delivery in section 3. Section 4 is broadly discussed the video content flows inside the hybrid architecture. Content blocking approximations at the proxy server is presented at section 5. Section 6 exhibits the content flow through the hybrid architecture. Section 7 is presented by the corresponding simulation with respect to proposition and algorithm and environment. The conclusion is presented in section 8.

## 2. CONTENT ENCODED MACHINISHM

The basic task within the transformation process is to reduce the spatial redundancy that always or already present in the residual. The transformation is done by orthogonal basis functions. Encoding technique is used to encode independently and corresponds to the sequence in the collected set of frames. The Joint Photographic Experts Group (JPEG) format is used to compress the still images for the set of images. The sequences of still images are arranged in increasing order, independently and individually. The MJPEG format is the Motion JPEG video compressed and encoded using JPEG. Each video signal that carries information, contain a definite amount of redundancy. So the video sequence contains redundancy. The aim of the video compression is to remove or minimize the redundancy presents in the video signal as much as possible. Multimedia Compression is a technique that has an enormous role in the digital multimedia application. Multimedia systems require efficient data compression methodology, mainly to reduce the number of storage servers. Specifically for those storage devices that are relatively slow, to transmit video data at real time. The video compression techniques successfully reduce storage device requirements. Digital data compression is done by using various computational algorithms and procedures. These are implemented either in software or in hardware. The compression techniques are classified according to the approaches, and implementation likes lossless and lossy. The Lossless approach can recover the original representation perfectly. It has been observed that the Lossy approach recovers the presentation with some loss of accuracy. However, the lossy techniques provide with higher compression ratios. Therefore, Lossy compression is more applied in video content compression than the Lossless approach. The compression technique is presented in Figure 1.

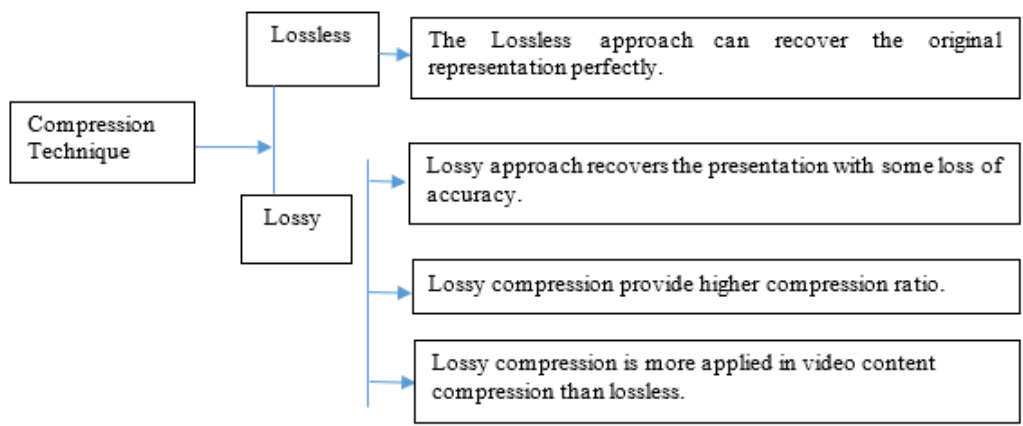

**Figure 1.** Compression of technique

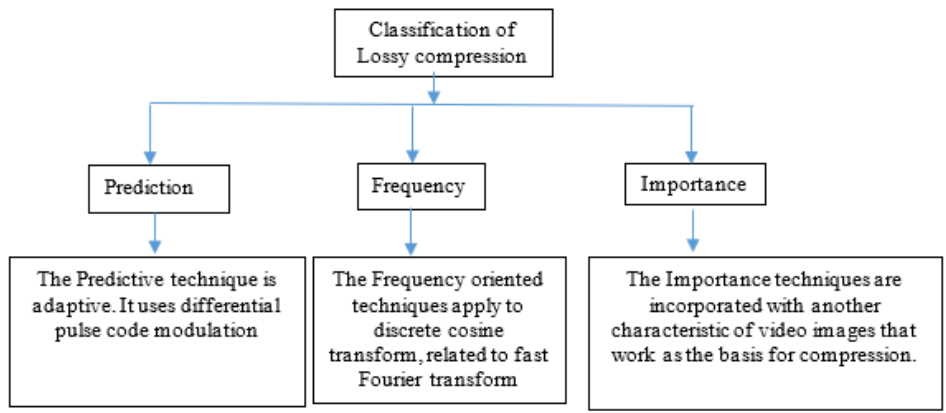

**Figure 2.** Classification of lossy compression

**Table 1.** Comparison of lossy and lossless technique

|  | Advantage | Disadvantage | Used file type |
|---|---|---|---|
| **Lossy** | It uses very small file sizes. A number of tools, plugins, and software support it. The file takes less space. Files can be transmitted very quickly. | Quality degrades with a higher ratio of compression. Can't get original back after compressing. Lose quality, Lose resolution, Cannot restore to original state. | MPEG Audio Layer 3, Moving Picture Experts Group Phase 3 (MPEG3), MPEG player III (MP3), Moving Picture Experts Group Phase (MPEG), Joint Photographic Experts Group (JPEG). |
| **Lossless** | No loss of quality, slight decreases in image file sizes. | Larger files use for lossy compression. The file takes up a lot more space. Files cannot be transmitted as quickly. | RLE (Run Length Encoded Bitmap), LV (Flash video), Zip file (archive file). |

The lossy technique is further classified into prediction, frequency, importance based on the techniques and implementations. It is presented in Figure 2. For example, predictive technique is adaptive differential pulse code modulation. It predicts subsequent values by observing the previous values. The frequency oriented techniques apply to the discrete cosine transformation (DCT). It is related to fast Fourier transforms. These Important techniques are incorporated with other characteristics of video images that work as the basis for compression. The DVI (Digital video interactive) technique employs on color lookup tables and data filtering technique. It is an example of Importance- oriented technique.

The other type of compression technique is Hybrid compression techniques. It is a combination of several approaches such as DCT and vector quantization or differential pulse code modulation. Various groups have already set their standards for digital multimedia compression technique, based on JPEG, MPEG, and px64 etc. In most video compression standards such as JPEG, MPEG, and H.26X technique, that compressions are based on 8x8 DCT (discrete cosine transform) to exploit the spatial redundancy. The motion compensation technique is incorporated to exploit the temporal redundancy present in the video signal. The scrambling procedures of the video content are applied to the coefficients, after the quantization stage. The motion vector information is collected during the motion estimation stage. All the video compression standards are followed by a common hierarchical structure platform. The video sequences are segmented into groups of pictures (GOP). Each GOP is composed of an intra-coded frame (I frame) followed by some forward predictive coded frames (P frames) and bi-directional predictive coded frames (B frames). Intra frame prediction is a prediction technique that considers the reference blocks from its own frame. It exploits the spatial correlation between blocks. These do not refer to other frames. There exist three important prediction modes for intra coding. Those are Intra 4x4 mode, Intra 8x8 mode, and intra 16x16 mode.

The Intra 4x4 mod is presented by 4x4 sized blocks. It is predicted by the spatial neighboring pixel. Intra 8x8 mode works only under the FRExt (fidelity range extensions) with high profiles. Comparing with other modes, it allows 8x8 block size, which generally provides more flexibility for HD (High definition) quality video. The intra 16x16 mode has four modes such that those are allowed for 16x16 sized macro blocks. The first three modes work similarly as the corresponding modes in the Intra 4x4 model. The fourth mode is a planer prediction about the samples.

The lowest layer is the 8x8 block, which are the units for DCT transformed coding. A '2x2' square of luminance blocks, together with the association of chrominance blocks, form a macroblock. It is the unit for quantizes, selection and motion compensation. A horizontal strip of contiguous macroblocks forms a slice. A frame consists of slices. Each layer except the block layer has a header followed by the data at lower layers. In 8x8 DCT transforms coding, the 64 transformed coefficients are in a Zig-Zag orientation such that coefficients are arranged approximately in order of increasing frequency, reference Figure 3. The goal of encoding is to represent the coefficients by using low bits as possible. This is accomplished in two steps. Run length coding (RLC), which is used at the first level of compression. Variable length coding (VLC) is carried in the next level of compression. H.264/MPEG4-AVC (Advanced Video Coding) is the latest video coding standard of the ITU-T Video Coding Experts Group (VCEG) and the ISO/IEC, Moving Picture Experts Group (MPEG). The MPEG-4 is a collection of coding tools and maintains a simple profile. The most current implementations for temporal scalability are video coding H.263, MPEG-4, and secular temporal sub-band coding.

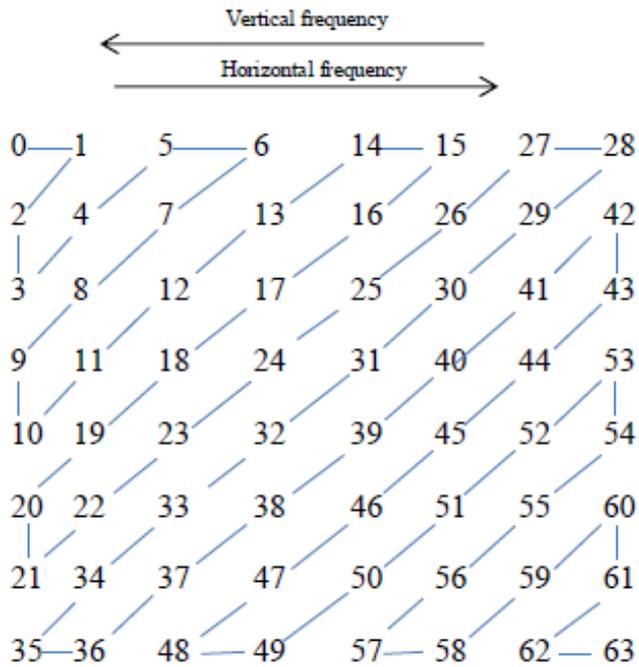

**Figure 3.** Zig-Zag ordering of DCT coefficients

These coding schemes adopt a linear uniform sampling scheme that disregards the varying semantic importance of different frames or segments. H.264/MPEG4-AVC has recently become the most popular and widely used video coding standard. The deployment of MPEG2 has occurred at the dawn of digital television. Now, H.264/MPEG4-AVC rapidly overtakes MPEG2 in common areas of applications. It covers all common areas of video applications such as mobile services, video-conferencing, IPTV, HDTV, and HD video storage. The latest video coding is H.265/HEVC, which is the successor to H.264.H.265 is meant to double the compression rates compared to H.264. This coding format allows the propagation of 4K and 8K video content objects over the existing network systems. With the rapid improvement of network protocol, multimedia technology has made video-on-demand more and more popular in our daily life. The daily life applications like mobile services, video-conferencing, IPTV, HDTV, and HD video storage, H.265/HEVC coding is highly implemented [22].

## 3. TRANSCODING FOR VIDEO STREAMING

The primary 'video storage servers store raw material' in pre-compressed or in pre-compiled format and perform quality adaption on the fly to simulate a real-time encoder as much to reduce the complexity. The implementation of transcoding correctly needs to decode video content fully. The reconstructed sequence is obtained by re-encoding the sequence with the new set of coding parameters by cascading. This approach also includes the motion estimation module. It is likely to possess maximum complexity. Various numbers of architecture are proposed to improve the cascade transcoder [22-26]. Those architectures are considered by removing the motion estimation module. Some of them are as follows.

### 3.1 Cascade decoder and encoder architecture

Cascade Video transcoder is the simplest transcoder. This architecture takes source bit stream that is decoded by Variable Length Decoder (VLD).

### 3.2 Open loop architecture

The Open- Loop Architecture is associated at the decoder end. Source Bitstream submits to VLD, which produces the DCT coefficient. The inverse quantization is preceded on the collected DCT coefficient at the encoder end. The DCT coefficients are again re-encoded by using different quantizer (Qt). Now, it is passed through the Variable Length Coding (VLC). It is computationally efficient and fast with less computational complexity. It has been existing drift error.

### 3.3 Close loop architecture

Closed Loop transcoder contains a feedback loop. The closed loop architecture has the ability to remove the mismatch between the residual and the predicted frame. The major difference between the closed-loop architecture with the cascaded decoder-encoder is in the area of the reconstruction loop. It is operated in the pixel domain. This architecture has only one DCT/IDCT pair.

### 3.4 Frequency domain transcoding

This architecture is used in Transcoder by re-encoding the DCT frequency coefficients. The frequency domain transcoding architecture works at the decoder end. To get the DCT coefficient first, it passes the source bit stream (Vs.) to Variable Length Coding (VLD). Then the inverse quantization is performed. At the encoder end, the encode motion compensation is done by applying re-quantization and VLC. Now, the DCT values are stored in the reference frame (R). It requires less computation with existing drift error and lacks flexibility. The above four architectures are simplified cascaded transcoder.

To enhance the methodology, it is required to eliminate the modules progressively with the expense of drift or coding. The little modification on DCT module at the encoder loop is moved before the summation process, with consideration such as a process is linear. The relocate DCT and the IDTC module at the decoder are canceled out to each other. Therefore, the IDCT module at the decoder loop is removed at the current stage. The last simplification is to merge the two frames (both the encoding loop and decoding loop). Here, the motion compensation is done only for once. With the help of those simplifications, spatial domain transcoder (SSDT) is developed. Figure 4 presents the SSDT. The comparison of different architectures is presented in table 2.

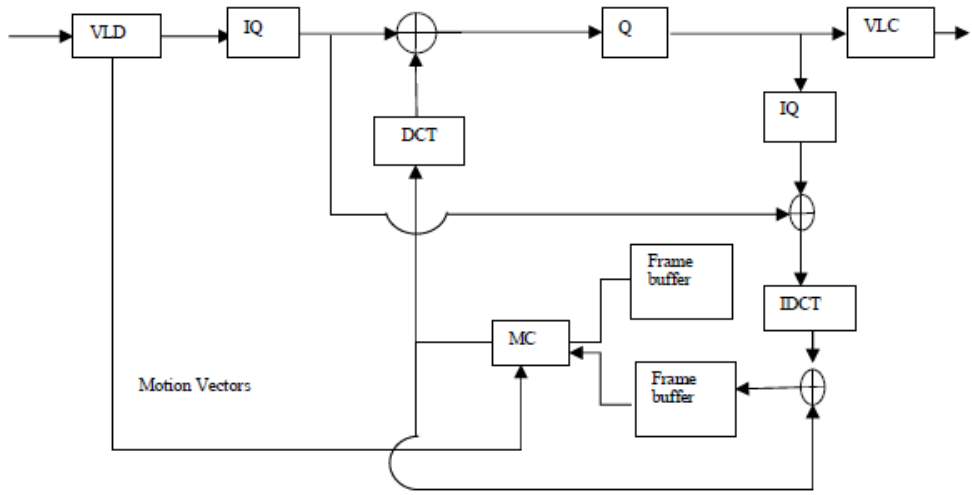

**DCT:** Discrete Cosine Transform
**IDCT:** Inverse Discrete Cosine Transform
**Q:** Quantization MV: Motion Vector
**IQ:** Inverse Quantization
**F:** Frame Memory
**MC:** Motion Compensation

**Figure 4.** Spatial domain transcoder

**Table 2.** Comparison of different architecture

| Different Architecture | Advantage | Disadvantage |
| --- | --- | --- |
| CASCADE DECODER AND ENCODER | It is a simple and flexible architecture. This architecture reduces the computational complexity. In this architecture the reconstruction loop operating in the pixel domain. | This architecture introduces drift error. |
| OPEN LOOP | It is computationally efficient. This architecture processes the data fast. It requires minimum transcoding complexity. | This architecture has been existing drift error. |
| CLOSE LOOP | This architecture has the ability to remove the mismatch between the residual and the predicted frame. | This architecture has only one DCT/IDCT pair. |
| FREQUENCY DOMAIN TRANSCODING | It needs less computation. | This architecture lacks flexibility. |

## 4. CONTENT FLOW INSIDE THE HYBRID ARCHITECTURE

The transcoder maintains a visual quality standard to suit the target viewers. The characteristics of content transportation also require to consider transcoding to maintain the quality of service (QoS) at the viewer end. Therefore, the smooth transportation of video content over the heterogeneous network is becoming more challenging. Lots of literature addresses this issue with various concepts based on content driven network (CDN), pure Peer-to-Peer network architecture, and hybrid as a mixed of CDN and Peer-to-Peer network architecture [8, 27]. However, the design of an efficient P2P-VOD system with high peer bandwidth utilization with low maintenance cost remains a challenging issue. The demand for smooth video content transfer over the low bit-rate channel is increasing at a fast pace. Some real-life applications are newscast, video conferencing, distance learning, entertainment, etc. So, some sort of traffic flow mechanism and congestion control mechanism with the help of admission control at the switches and junction point has to be considered for the enhancement of content transportation.

It is required that the packet drop at switch and junction point needs to be considered as one of the major characteristics of video content flow. However, a major bottleneck is observed during the timely delivery of a large amount of video content through the limited bandwidth links. The issue that stands here is how efficiently and perfectly the overlay network adapts dynamically formed virtual clusters of peer nodes.

The video content is pushed from the origin storage i.e. primary storage to the peer nodes over the tree-shaped overlay network. The tree-based overly network is addressed by multi-tree streaming approach [28-30]. Here the compact topology is dynamically changed with the multiple sub-trees in place of one single tree [5].

However, the mash based content delivery approach gives better results for both video on demand and for the live video streaming.

The letter concept gives better performance observed in PPlive, and Livesky application [2, 4]. To achieve the quality of service, video content delivery delays are to be handled smoothly. The delay occurs due to the search for video content. Delay is also occurred due to the propagation of requests for content searching and transportation of content video from the neighboring peer nodes.

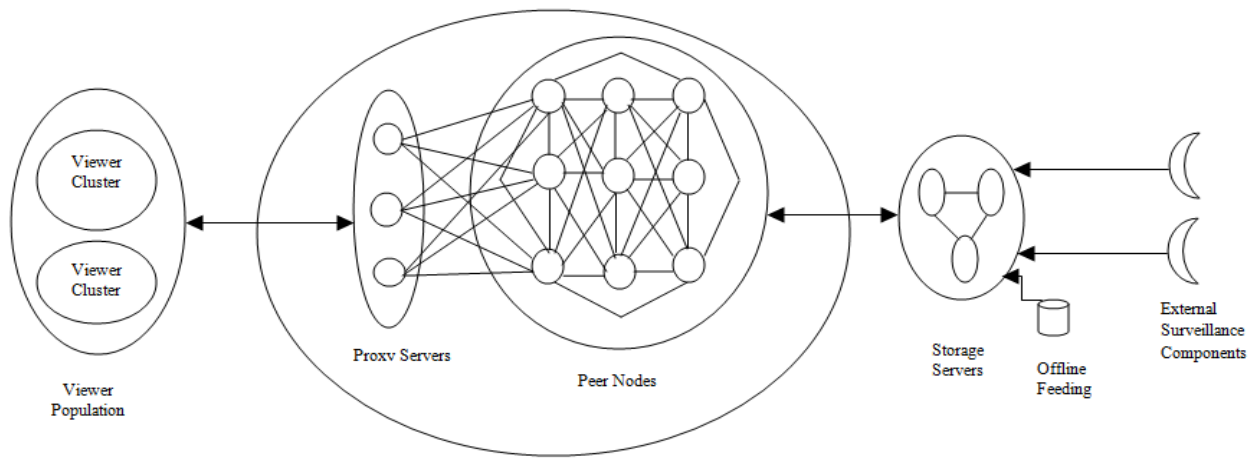

**Figure 5.** Compact hybrid architecture for video on demand system

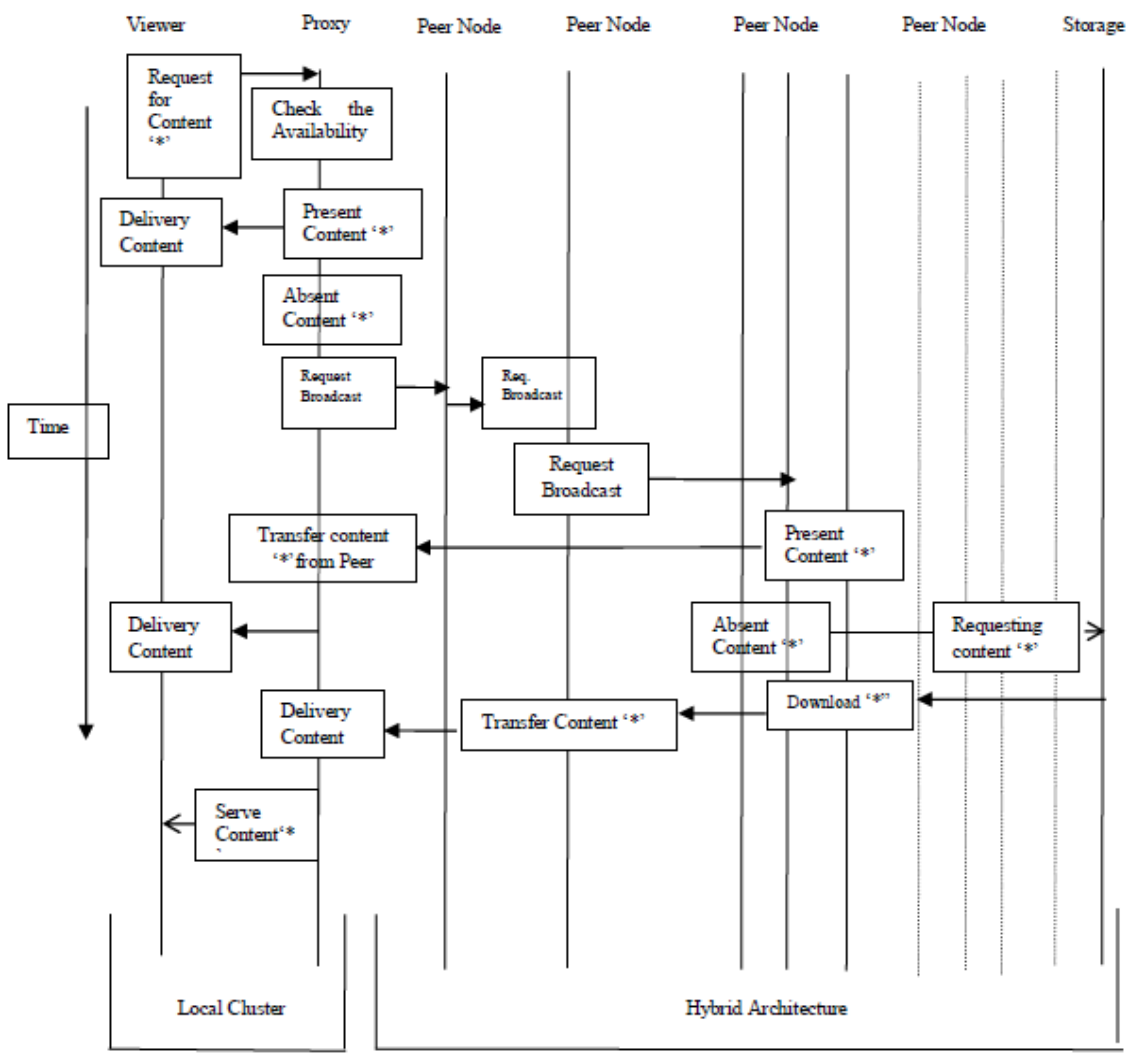

**Figure 6.** Content transfer working mechanism

So the quality of service mainly depends sharply on fast and correct storage finding with considering transcoding. To find the correct video stream by using minimum hop count towards the peer node requires to be addressed. During content transportation through the heterogeneous network, the aggregate stationary bit rate does not exceed the maximum threshold level from the viewer perspective. Here, it is required to handle the transcoding efficiently. Figure 5 presents the physical configuration of the overall structure of the compact hybrid architecture. Here the clusters of viewers are presented by the symbolic viewer node. The proxy servers are connected to the peer nodes of multitier mesh architecture at the very next level of hybrid architecture references Figure 5. The proxy server stays at the height level in reverse order of hierarchy within the hybrid architecture. The proxy servers are not directly connected to each other in the hybrid architecture. They are connected to each other through the next level of peer nodes.

So, from the proxy level up to the lowest level i.e. the primary (storage server container) forms a compact architecture. The proxy server at the height level usually imports the video content from the next level of peer nodes. The video content is transferred according to the flow diagram, presented in Figure 6. There is an option in hybrid architecture to store the copy of the live stream through the connected 'surveillance components' for real-time object tracking reference Figure 3. In this architecture, the live stream is archived to storage servers. The video content is directly imported to the proxy server, partially or completely according to the content available to the peer or to import from storage inside the hybrid architecture. The architecture is composed of local viewer clusters along with peer nodes and primary storages. The proxy server is a virtual part of the local viewer cluster as well as the hybrid architecture. Peer nodes belong to different levels in a hybrid architecture. The last location is fixed for the primary video storage server. The initial request tries to retrieve content from the local proxy server; either the content is available there or not. If the local proxy server cannot serve the requested content, then the request is broadcast to the neighboring virtual cluster of peer nodes at the next level of hybrid architecture. Eventually, the request is propagated throughout the architecture until it retrieves the required content. During propagation of request messages, if any virtual cluster of peer node has the required video content, then that peer node rebroadcasts the availability of content. The content is transferred according to algorithm 1. The procedure is described in Figure 6. The obtained virtual cluster of peer nodes at real time is followed according to Figure 7. The video data forward mechanism is one type of Mesh push from the sender peer node, and Mesh pulls from the receiver peer. The copy of the content video is stored at the local proxy server before delivery to the local viewer cluster. The traffic flow mechanism is classified into two main sections. In the first section, we consider the enormous amount submitted requests processed at the proxy server, with the consideration, that the requested video is present at the proxy server. The video content is directly streamed to the viewer. In the second section, we consider the viewer request is blocked at the proxy server. The blockage occurs due to unavailability of the video content at the proxy server. We consider in this situation, whether the video is partly presented or not, at the proxy server. So, the remaining portion of the video content is partly or completely imported from the peer node or from the last level of the hybrid architecture i.e., archived server.

## 5. VIDEO CONTENT BLOCKING APPROXIMATIONS AT THE PROXY SERVER

Let us consider maximum $n$ numbers of requests are served by the proxy server at any instance of time. If $R_d$ is the overall disk bandwidth at the proxy server and $R_p$ is the client request playback rate of the viewers (including all types of the interactive operation) then at the proxy server, $n \leq \left\lfloor \frac{R_d}{R_p} \right\rfloor \leq \eta$. Here, $\eta$ is the considered admission control threshold in the case of maximum $n$ requests served by the proxy server [1]. Let the proxy server has a total $C$ number of ports. Here, we considered $k$ classes of service requests (including all types of the interactive operation) approached at the proxy server with $\lambda_i$ (Poisson rate), where $i = 1$ to $k$. The proxy server ports capacity are divided into partitions with each partition allocated ports capacity $C_j$ for $j = 1$ to $k$ and the server ports occupancy for class $i$ type of service request is $Q_i$. Viewer request comes for the $i^{th}$ type of service. If it is served by the $j^{th}$ block of the port partition at the proxy, serves then $1 < j$. Here, $B_i B_{i+1} \cdots B_{j-1}$ are the events that all the partition at the proxy server are blocked from $i$ to $j-1$, the only $j^{th}$ partition has at least one available free port that is presented by the event $A_j$.

Now,

$$p(A_j / B_i B_{i+1} \cdots\cdots B_{j-1})$$
$$= \frac{p(A_j B_i B_{i+1} \cdots\cdots B_{j-1})}{p(B_i B_{i+1} \cdots\cdots B_{j-1})}$$
$$= \frac{p(A_j) p(B_i) p(B_{i+1}) \cdots\cdots p(B_{j-1})}{p(B_i) p(B_{i+1}) \cdots\cdots p(B_{j-1})}$$
$$= p(A_j) \qquad (1)$$

By sequential search for free the port at the proxy server (partitions wise) and by using the geometric distribution with parameter $\left(\frac{1}{k}\right)$. we get the following expression,

$$p(A_j / B_i B_{i+1} \cdots\cdots B_{j-1}) = \left(1 - \frac{1}{k}\right)^{j-1} \left(\frac{1}{k}\right) \left(\frac{C_j - Q_i^j}{C_j}\right)$$
with $j = 1, 2, 3, \ldots$

The blocking probability from $i^{th}$ partition to $j^{th}$ partition is represented as

$$p_b(B_i B_{i+1} \cdots\cdots B_{j-1}) = p_b(B_i) p_b(B_{i+1}) \cdots\cdots p_b(B_{j-1}) \qquad (2)$$

Since $p_b(B_m \cap B_n) = \varnothing, \forall (m,n) \in I$ and $m \neq n$
Where $i \leq m \leq j-1$, $i \leq n \leq j-1$
Eq. (2) express as with Erlang,

$$p_b(B_i B_{i+1} \cdots\cdots B_{j-1}) = \left[\frac{\frac{E_i^{C_i}}{|(C_i)|}}{\sum_{k=0}^{C_i} \frac{E_i^k}{|(k)|}}\right] \left[\frac{\frac{E_{i+1}^{C_{i+1}}}{|(C_{i+1})|}}{\sum_{k=0}^{C_{i+1}} \frac{E_{i+1}^k}{|(k)|}}\right] \cdots\cdots \left[\frac{\frac{E_{j-1}^{C_{j-1}}}{|(C_{j-1})|}}{\sum_{k=0}^{C_{j-1}} \frac{E_{j-1}^k}{|(k)|}}\right] \qquad (3)$$

If $h_i$ is the port occupied time by the $i^{th}$ type of service request from viewer cluster and $T_n$ is the total time that each port at the $n^{th}$ partition is busy, then the corresponding Erlang is expressed as

$$E_n = \frac{1}{T_n} \left[\sum_{i=1}^{C_n} \lambda_i h_i\right] \qquad (4)$$

## 6. CONTENT FLOW THROUGH THE HYBRID ARCHITECTURE

In this case, the proxy server has the required content as partly or may not possess a complete portion of the content. The situation needs that the content video is partly or completely imported from the peer node or from the archiving server. Here we consider $n$ numbers of active links that simultaneously transfer the content from the peer nodes via proxy servers over the hybrid architecture. Let $X_i$, $i=1,....,n$ are the random variables, which present the content transferring. Now, $E(X_i) = \mu_i$, $i=1,....,n$ are the means of transferring rate for an individual ($i^{th}$ viewer) request. It is assumed that $X_i$, $i=1,....,n$ are the mutually independent random variables. So the aggregate content transfer at that session is expressed by $B = E(\sum_{i=1}^{n} X_i)$. According to the proposition 6.1.1, we conclude that, $A_p \leq E(\sum_{i=1}^{n} X_i) \leq A_P^*$ for $p \geq 2$, where $A_p$, $A_p^*$ are real constant in ($p$ - norm) for finite dimensional vector spaces.

### 6.1 Proposition

If $x_1, x_2, x_3, \cdots, x_N \in L^p$ are the random bit rate used by the peer nodes inside the hybrid architecture at any session, then there exit bonds for $E(\sum_{i=1}^{N} x_i)$ with positive constants $c_p$ and $C_p^*$ such that, $c_p(\sum_{i=1}^{N} x_i^2)^{\frac{1}{2}} \leq (E \sum_{i=1}^{N} x_i) \leq C_p^*(\sum_{i=1}^{N} x_i^2)^{\frac{1}{2}}$, $L^p$ is the Lebesgue space.

*Prof:*

If $\{z_i\}_{i=1}^{N}$ are independent identical random variables with probability $p(z_n = \{0,1\}) = \frac{1}{2}$ subject to channel link are active or inactive. According to Khintchine inequality in the $p$-norm for finite dimensional vector spaces, for $0 \prec p \prec \infty$. There exist some constant $c_p \succ 0$, $C_p^* \succ 0$ such that $c_p(\sum_{i=1}^{N} x_i^2)^{\frac{1}{2}} \leq (E \sum_{i=1}^{N} x_i) \leq C_p^*(\sum_{i=1}^{N} x_i^2)^{\frac{1}{2}}$. This is holding true for a finite number of active channels links.

Usually, the values of $c_p$ and $C_p^*$ depends on $p$. So for the $p \geq 1$, the $p$-norm for $x = \{x_1, x_2, x_3, \cdots, x_N\}$ is expressed as $\|x\|_p = (|x_1|^p + |x_2|^p + \cdots + |x_N|^p)^{\frac{1}{p}}$

*End of proof* ∎

The above proposition 6.1.1 holds true if there exist the least number of active links $k^*$ (say) for all $n \succ k^*$, n is the possible number of all channels. Let $\psi_{(s)}^{\dagger}$ be the equivalent capacity of the network, the minimum required bandwidth to ensure the aggregate stationary bit rate. The aggregate stationary bit rate is required to approximate the transcoding as discusses in section 3. Aggregate bit rate maintains the video resolution into or within a certain standard. As all the individual and shared links are used by the maximum bandwidth. The proposition 6.1.1 exhibits the upper bond content i.e., $Max \sum_{i=1}^{k^*} E(X_i) \leq \psi_{(s)}^{\dagger}$, here $\psi_{(s)}^{\dagger}$ is the capacity of the network. It maintains that the aggregate stationary bit rate streaming remains at the same value for the short burst period of time. Without the loss of generality, all the above inequality holds true for the smooth content transfer through the hybrid architecture. The channel distribution can be obtained through Markovian model. For every viewer node, there exists at least one active channel, including the case of broadcasting or multicasting of video data stream from the Proxy server [31-32]. At any level of multi-tier architecture, some of the peer nodes content the full portion of contain or part of the required content. For continuous streaming, it is required that the video content is fully available to the proxy server. In a real scenario, the large number of viewers can concurrently decode the video within the range of bits rate 400 to 800 Kbps [33]. Dynamic Page replacement policy at the proxy server has a major role in the performance issue in the VOD system [34]. The literature [5, 34] showed that proper cache memory handling can reduce the traffic load to some extent. The simultaneous viewers basically exhibit asynchronous tendency in nature. Clearly, the viewer desires to access any part of that video at any instance of time. Due to the playback option at the viewer end, the VOD system needs enough number of sequential video content available to the media player buffer of the viewers. In general, the viewer assesses that video through the proxy server. The video objects (a collection of video frames) are fed to the media player after the sequential arrangement of frames at viewer buffer. If the complete video is not available to the proxy server, then the unavailable portions of the content are imported by the proposed algorithm 1 and process flow is presented in Figure 6. It inhibits the multi-channel proxy server from any number of peers at any level of the hybrid architecture. The proxy server is the part of the mesh architecture which is placed at the maximum level of the multi-tier architecture. The proxy servers are not the peer servers, so there exist no interlinks between the proxy servers. This is considered for the requirement of security purpose. The billing servers are installed at the level $l_1$ i.e., at level one in multitier architecture (which looks like an inverted tree). The level zero i.e., $l_0$ is the position for an archive storage server. The proxy cache memory is updated or refreshed by the 'least recently frequently used' concepts. Here, $l_0$ is the level for archive storage server and depth level (for the higher value of $d$) i.e., $l_{d+1}$ is the position for cache proxy server. Let $\lambda$ is the average fraction of full streaming at a session that each of the peer nodes shares or contribute during that session clearly, $0 \prec \lambda \prec 1$. $P(l)$ be the sum of the streaming capacity at level $(l)$. Now $N(l)$ be the shearing capacity of the peer nodes at the level $(l)$. The steady traffic flow in a balanced state at any level $(l)$ depends on the video streaming receive from the peer nodes i.e., from the level $(l-1)$ in the multitier hybrid architecture. In a session of synchronous chunk transfer, the shearing capacity of the peer nodes at level $(l)$ is equivalent to the product of an average fraction of shearing capacity of the participating peer nodes at the level $(l-1)$.

This is expressed by $N(l) \approx \lambda N(l-1)$. so, $N(2) \approx \lambda N(1) = \lambda C_1$, as $C_1$ is the shearing capacity at level $l=1$. Since the level $l=0$ is for the archive storage server, no peer node exists at that level according to Figure 5. In real scenario, peer nodes at the level $(l)$ get chunk video objects when the level $(l-1)$ participating peer nodes send the required video objects for them. The throughput of the system is contributed by the average fraction of full streaming of participating pees nodes. Some portion of the bandwidth is always used for chunk transfer or exchange among the peer nodes belong to the same level according to Figure 7.

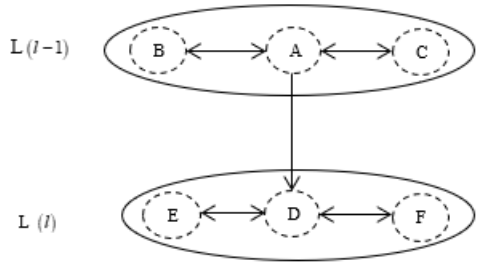

**Figure 7.** Virtual cluster formation

It proceeds to the expression

$$P(l-1) = P(l) + N(l)$$
$$\Rightarrow P(l) = P(l-1) - N(l)$$

Dynamic cluster connections are presented in Figure 7. It is observed that A, B, C are the peer nodes at the level $(l-1)$ and D, E, F, G are the peer nodes at the level $(l)$. The dynamic clustering is a logical cluster construction among the peer nodes at every level for real-time requirement at any session. The peer of nodes B and C at level $(l-1)$ exchange the chunk video objects, or transfer among themselves. D is the cluster head at a level $(l)$ that downloads video chunk from the dynamic cluster head 'A' from level $(l-1)$. The downloading chunk of video objects from the lower level, give priority over the 'chunk of video objective' transfers in the same level of multitier hybrid architecture in that session. From the previous expression, it can be preceded as follows:

$$P(l) = P(l-1) - N(l) \quad \text{for } 0 \prec \lambda \prec 1$$
$$= [P(l-2) - N(l-1)] - N(l)$$
$$= [P(l-3) - N(l-2)] - N(l-1) - N(l)$$
$$= \cdots$$
$$= \cdots$$
$$= \cdots$$
$$= P(0) - N(1) - N(2) - N(3) - \cdots - N(l)$$
$$\approx P(0) - C_1(1 + \lambda + \lambda^2 + \cdots + \lambda^{l-1})$$
$$\approx P(0) - C_1 \frac{1-\lambda^l}{1-\lambda}$$

$$P(l) \approx P(0) - C_1 \frac{1-\lambda^l}{1-\lambda}$$

The above expression represents the maximum streaming received at any level $(l)$. Which is clearly depends on how much video data is streaming from the level $(l=0)$, i.e., from archive storage. If the number of level increase then volume (amount) of received stream decrease since $(0 \prec \lambda \prec 1)$. Here, $p_{k_l}$ is the probability that at least $k^l$ out of $N_l$ peer nodes are actively participate at the $l_{l-1}$ level of the hybrid architecture. Active means that the peer has the required encoded portion of the video file in the content and during content transfer the end-to-end, links are live at that session.

In a descriptive form, we consider $Y_l$ as the event that at least $(k^l + 1)$ number of peer nodes are active at the level $l$, $Y_{l-1}$ is the event that at least $(k^{l-1} + 1)$ number of peer nodes are active at the level $l_{l-1}$ and so on.

**Algorithm 1. Heuristic Path Search Algorithm, initiate from the proxy server**

```
Var Q : Queue of Integer
Var v : node
Input    { L : Queue length
           Δt : start time
           t_s : Session period
           C : Collection of virtual cluster of peer nodes
         }
Initialization { Boolean matched_query_not_found ← True;
                 Length ( Q ) ← L ;
                 Hop_count = 0;
               }
while (t ≠ (t_s + Δt)) do
    {
begin
Q ← Enqueue (S) , such that S ⊂ C // S is the cluster head at each level
    while (matched_Quary_not_found) do
        { v ← dequeue (Q) // Extract from Queue
          If (query string) ∈ v
          then
          {    Hop_count = Hop_count +1;
               Output "count found"
               Output " Hop_count"
               match_query_not_found ← false
          }
          else
          for each (w) ∈ {Single hop virtual cluster}
          {
            if ( query string ∈ w )
            then
            {
              match_query_not_found ← false
              Hop_count=Hop_count+1;
              Output " Content transfer to proxy server";
              Output " Hop_count";
              Exit
            }
            else
            continue // search Query in other virtual cluster at same level
          } // end for each
        } // end while inner loop
        t ≠ (t_s + Δt)
    } // end outer while
    Output ("Content is not found");
End
```

Since,

$$p(Y_l \cap Y_{l-1} \cap \ldots \cap Y_0) \leq 1 \quad (5)$$

So, we get the expression

$$= \sum_{k=k^l+1}^{N_l} \binom{N_l}{k} \rho^k (1-\rho)^{N_l-k} \quad (6)$$

The aggregate content transfer depends upon the active participation of peers at the levels $l_l, l_{l-1}, \ldots, l_0$.

The distribution $B = E(\sum_{i=1}^{n} X_i)$ is obtained from the proposition (6.1.1). Since downloading of content at any level depends upon the upload capacity of the peer node at the just adjacent lower level in the hybrid architecture for every session. According to the expression (5) and (6), we get

$$\prod_{i=l}^{1} \sum_{k=k^i+1}^{N_i} \binom{N_i}{k} \rho^k (1-\rho)^{N_i-k} \leq 1 \qquad (7)$$

## 7. SIMULATION RESULTS AND DISCUSSION

Figure 8 represents the simulation result obtained by NS-2 simulator. We consider the variable incoming request pattern at the proxy server in the VOD system. The client request comes to the VOD system with various service requests. According to the table-1 simulation parameters, the plotted Figure 8 is presented as to how rapidly the incoming request grows with respect to time. Initially according to the real traffic scenario at the time 0.24 seconds, the system received 50 requests. At the time 0.36 seconds the system received 200 requests. Figure 8 represents a smaller snap of the overall simulation run. Figure 9 presents the comparative study of the system throughput at the proxy server for different service requests submitted by viewer cluster population. The 1st strip shows the respective throughput of the system for the smaller size of the population clusters at the proxy server. The 2nd strip presents the system throughput at the proxy server; here cluster of the population is greater than the previous one. The simulation results clearly indicate that if the size of the viewer inside the cluster increases, then the system performance gradually decreases. According to Figure 9, in a real scenario the size of the cluster inside the viewer population is exponentially increased. To decrease the blockage at the proxy server and simultaneously import the new video from the primary storage, the VOD system subsequently used the hybrid architecture (i.e. import video from peer nodes or from the primary storage). The rated output at the primary storage is maintained by transcoder. It is used to control and stabilize the content streaming. The transcoder maintains visual quality for the requested viewers. It is used with the help of ffmpeg command of Evalvid tool that is imported into NS2 simulator. The primary storage server contains the raw and compressed file. The YUV file is usually created by decoding the raw file. At first, it is converted to (*. mp4) file format. Then the mp4 container simultaneously creates a list of reference videos. The frames are packetized to transport over the hybrid architecture. The mp4 trace command is used to send the compressed (*.mp4) file to the requested viewer buffer through the hybrid architecture and it passes through the available port at the proxy server [5, 35-40]. The content of video (i.e. 'packetized frames') is transported by RTP/UDP protocol. This is implemented with the help of EvalVid-RA. It is a Tool-set, used for rate adaptive VBR (video bit rate) to handle transcoding efficiently. It is integrated with NS-2, with some modification in the EvalVid Version, 1.2 and in the ns-2 interfacing code. The average VBR rate controller is set to 600kbit/s for that session. The hybrid architecture, including the proxy server, has 15 levels. At the highest level $l_{14}$, proxy servers are installed. At the lowest level i.e., $l_0$ the primary storage servers are installed in the hybrid network. The levels $l_1$ up to $l_{13}$ may be used for the peer nodes in the hybrid architecture. Every level has 4 peer nodes that are always interconnected. The peer node belongs to the same level those forming the virtual cluster or exchange video content. The peer nodes at the higher level connected to the lower level peer node are used in unicast procedure i.e., only the higher level nodes can download the video content from the lower level peer nodes. In the simulation, every peer node maintains a dynamic connection. In a dynamic environment, every peer node maintains adjacent list size of one up to six. These links are dynamically selected by the peer node from the available 11 links assigned to that peer node. The 11 links are orientated as follows, 4 links are used for unicast downloading from the lower-level peer nodes. Three links are used for the forwarding (broadcasting) viewer requests to the higher-level peer nodes. The other 3 links are used for the content transfer at the same level of peers. The link capacity maintains within the speed of 400kbps to 800kbps. 400kbp is the average link capacity between the viewers to the proxy server. 800 kbps link capacity is maintained at the storage server to the next higher server. At the intermediate-level, peer nodes maintain the average link capacity 600kbps. The required parameters with the corresponding values and ranges are summarized in Table 3.

**Table 3.** Simulation parameters

| Parameter | Value/Range |
|---|---|
| Number of ports in each partition | 10 |
| Port access time | 120 sec |
| Number of partition | 20 |
| Number of level in Hybrid architecture | 15 |
| Number of Peers at each level | 4 |
| Raw Video frame rate | 30/sec |
| Group of picture length | 30 frames |
| Link Capacity | 400kbps to 800 kbps |
| Viewers connected to each of the Proxy | 20 to 40 |
| Simulation Time | 480 sec |

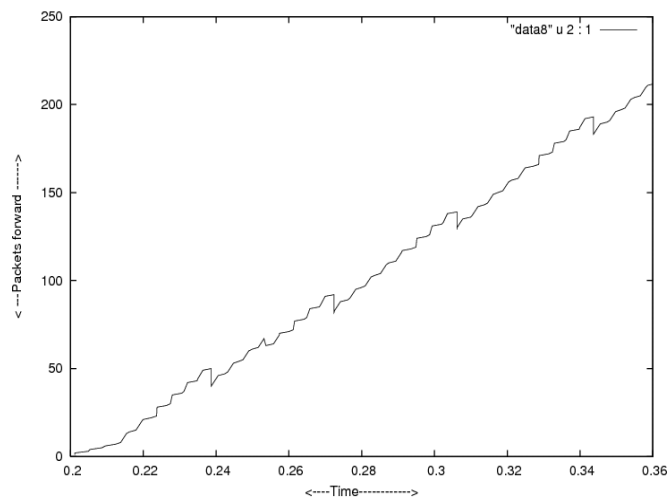

**Figure 8.** Incoming request pattern at the proxy

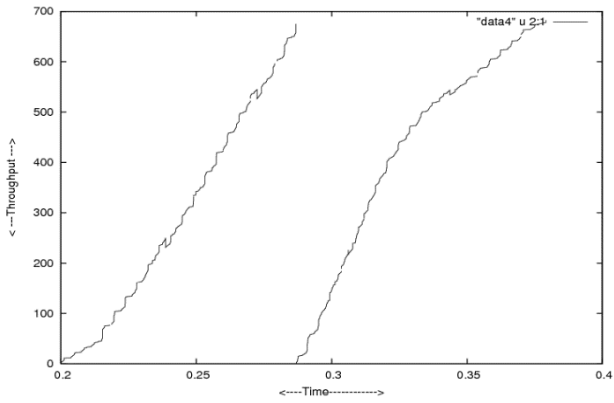

**Figure 9.** Throughput for two types of request from different viewer classes

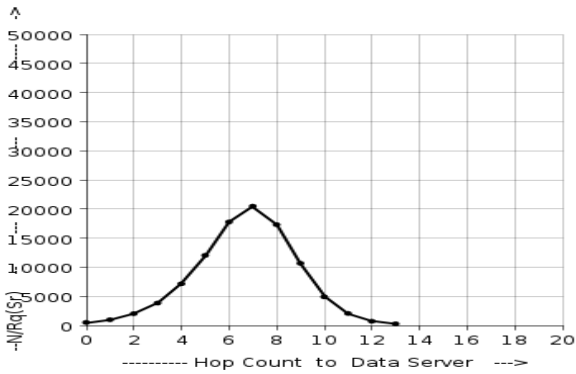

**Figure 10.** Single adjacent connection to each peer node

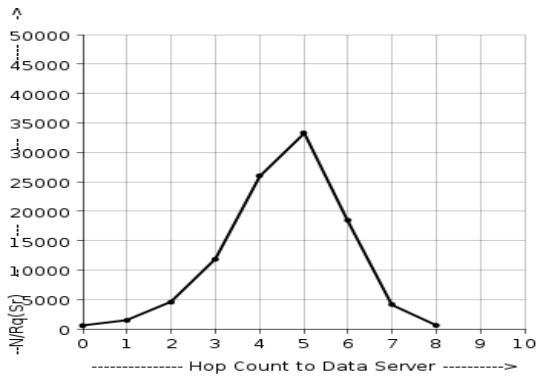

**Figure 11.** Two adjacent connections to each peer node

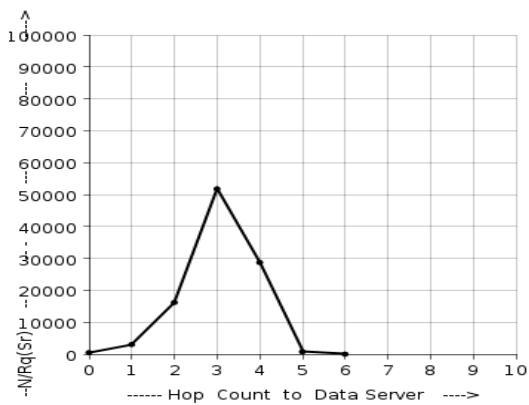

**Figure 12.** Five adjacent connections to each peer node

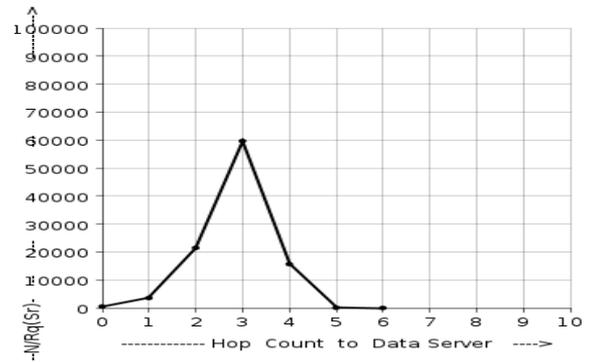

**Figure 13.** Six adjacent connections to each peer node

The simulation results are based on the request initiated by the proxy server and the proposed algorithm (1). The simulation results present in Figure 10, Figure 11, Figure 12, Figure 13, are snaps of the content flow when the proxy server does not possess the required content. In the plotted Figure 10 to Figure 13, the vertical axis present with the number of requests sent from the proxy server (cumulative) and the horizontal axis is the number of hops to that content storage. The number of trials, request initiative from the proxy server is approximately varied within (138000 and 174000). Figure 10 and Figure 13 present the simulation results of the minimum hop counts to import the required video content from the peer nodes or primary storage. The peer node at the level $i$, in the hybrid architecture, for $1 \leq i \leq 13$ maintains dynamic adjacent list size of 1 up to 6. If each peer node at every level maintains adjacent list size one, then the traffic flow within hybrid architecture follows as Gaussian like distribution, that is presented in Figure 10.

## 8. CONCLUSION

This work presented content blocking approximations at the proxy server to enhance the content flow in the video on demand system. If the blocking occurred at the proxy server, then traffic flows are carried inside the Hybrid architecture. The architecture is composed of Peer-to-Peer network with mesh shaped orientation of nodes. It clearly enhances the content delivery rate and improves the time elapse during content transportation to viewers. This work exhibits the improvement over the previous work on the area of content transfer, with respect to hop count, which explicitly reduces the delay. The huge number of simultaneous requests generated at viewer population makes congestion in the video on demand system. The Hybrid architecture and proposed methodology enhance the content transfer through the VOD system. The presented methodology effectively handles the huge traffic rate that approaches the proxy server. The proposed mechanisms smoothly manage the video data served by the distributed peer nodes belong to various levels of the hybrid architecture. The set of proxy servers put into a separate layer gives more robustness with respect to security in video on demand system